\documentclass[useAMS,usenatbib]{mnras}
\usepackage{amsmath}
\usepackage{amssymb}
\usepackage{graphicx}
\graphicspath{{./figures/}}
\usepackage{hyperref}
\usepackage{pdfpages}
\usepackage[authoryear]{natbib}

\newcommand{\eq}[1]{\begin{equation}\begin{split}#1\end{split}\end{equation}}

\newcommand{\eal}[1]{\begin{align}#1\end{align}}
\defcitealias{SPL94}{SPL}

\begin{document}
\title[Linear theory of tidally excited density waves]{The linear theory of tidally excited spiral density waves:\\application to CV and circumplanetary disks}
\author[Xu \& Goodman]{Wenrui Xu, Jeremy Goodman\\
Department of Astrophysical Sciences, Princeton University, Princeton, NJ 08544, USA
}
\maketitle

\begin{abstract}
  We revisit linear tidal excitation of spiral density waves in the disks of cataclysmic variables (CVs), focusing on
  scalings with orbital Mach number in order to bridge the gap between numerical simulations and real systems.  If an
  inner Lindblad resonance (ILR) lies within the disk, ingoing waves are robustly excited, and the angular-momentum flux
  they carry is independent of Mach number.  But in most CVs, the ILR lies outside the disk.  The wave flux and its
  scaling with Mach number are then very sensitive to conditions near the disk edge.  If the temperature and sound speed
  vanish there, excitation tends to be exponentially suppressed.  If the Mach number remains finite in the outer parts
  but the radial and vertical density scale lengths become comparable due to subkeplerian rotation, resonance can occur
  with acoustic-cutoff and stratification frequencies. These previously neglected resonances excite waves, but the
  Mach-number scaling remains very steep if the radial scale length decreases gradually.  The scaling can be
  less strong---algebraic rather than exponential---if there are sharp changes in surface density at finite sound speed.
  Shocks excited by streamline-crossing or by the impact of the stream from the companion are unlikely to be important
  for the angular-momentum budget, at least in quiescence.  Our results may also apply to circumplanetary disks, where
  Mach numbers are likely lower than in CVs.

\end{abstract}

\section{Introduction}
Since \citet{SMH86}, many numerical simulations of Roche-lobe overflow in semidetached binaries have
shown spiral shocks in the disk around the primary (e.g.,
\citealt{S-CR88,SM92,RS93,GLL98,Stehle99,Blondin00}).  \citet[hereafter SPL]{SPL94} interpret the
excitation of these shocks as a linear effect of the tidal field of the secondary and claim
reasonable agreement with their own nonlinear two-dimensional simulations as regards the torque
exerted on the disk.  Most recently, \citet{JSZ16,JSZ17} report spiral shocks in both hydro and MHD
simulations, and they interpret their results partly by reference to the linear framework of \citetalias{SPL94}.
 
According to standard linear theory for very thin disks (\citealt{GT79}), excitation of density
waves by a smooth periodic potential occurs only at resonances where the radial WKBJ wavenumber vanishes.  In
the two-dimensional approximation adopted by many of the works cited above, the only relevant
resonance is the inner Lindblad resonance (hereafter ILR), although others can occur in three
dimensions \citep{Lubow1981}.  \citetalias{SPL94}'s linear analysis considers a disk that is truncated well inside
the ILR and neglects the impact of the tidal stream (an intrinsically nonlinear process).  \citetalias{SPL94}
attribute the excitation they calculate to ``wings'' of the ILR.  However, \citetalias{SPL94}'s simulations, like
most others, adopt a smaller ratio of orbital to sound speed (hereafter \emph{Mach number}
\(\mathcal{M}=r\Omega/c\)) than is likely to obtain in most CV disks.  The simulations are not
sufficiently extensive to pin down the scaling of the wave torque with \(\mathcal{M}\) and
$r_{\max}/D$, the latter being the ratio of the disk's outer radius to the binary's separation.

These spiral waves are of more than technical interest.  During dwarf-nova outbursts,
angular-momentum transport and accretion in CV disks are probably dominated by magnetorotational
turbulence.  During quiescence, however, the disks are likely too cool and resistive for such
turbulence to sustain itself \citep{GM98,SLDF17}, whence another mechanism is required to drive
accretion in quiescence---if quiescent disks indeed accrete. An observed correlation between the
inferred viscosity parameter $\alpha_{\rm cool}$ in quiescence and the binary period or mass ratio
(these two are correlated because the main-sequence companion fills its Roche lobe) suggests that
tidal perturbations to the disk, and perhaps specifically tidally-induced spiral shocks, may be
involved \citep{Menou00,CSWSH12}.  While the disk expands during outburst, it appears to
shrink well inside the ILR in quiescence \citep{Smak71,Stanishev04,BBO16}.

Spiral features have been found by doppler tomography of some CV disks, notably IP~Peg and U~Gem
\citep{SHH97,NB98,Groot01,BMHMS05}.  The pitch angles of these observed spirals are too large for density waves at
realistic values of $\mathcal{M}$, however, and it has been suggested that the spiral features are due instead to the
vertical resonance \citep{Ogilvie2002a}, or simply to the nonaxisymmetry of the tidally induced velocity field in the
outer disk: i.e., the compression of closed orbits (neglecting pressure) along the line between the stars
\citep{Smak01}.

Accretion may also be driven by spiral shocks in the disks surrounding nascent jovian planets \citep{ZJS16}.  Here the
Mach number is expected to be lower than in CVs, making both the excitation (by the stellar tide) and the propagation of
these shocks more efficient.

The present paper focuses on the linear excitation of spiral density waves by the tidal potential, especially when there
is no ILR within the disk and the Mach number large.  We adopt idealized axisymmetric disk models with adjustable outer
radii and Mach numbers, and various density profiles.  (In models where \(c\to0\) at the disk edge, \(\mathcal{M}\)
pertains to somewhat smaller radii where conditions are more characteristic of the disk as a whole.)  This reduces the
excitation to a well-defined mathematical problem in linear ordinary differential equations, which we treat both
analytically and numerically.

This paper is organized as follows. In \S\ref{sec:waves} we review the equations governing linear density-wave
perturbations in an axisymmetric disk, their boundary conditions, and methods for calculating their excitation by a
tidal potential.  Section \ref{sec:polytropic_disk} analyses a model in which the sound speed tends smoothly to zero at
the edge, rather like an Emden polytrope with a free boundary.  It is shown both numerically and analytically that when
such a disk contains no ILR, linear wave excitation tends to be exponentially small at large \(\mathcal{M}\).
\S\ref{sec:resonances} examines cases in which there is a finite temperature and sound speed in the outer disk, if only
because of irradiation by the white dwarf and its companion, but the radial scale lengths of density and pressure become
small due to subkeplerian rotation: the stream through the inner Lagrange point arrives with constant specific angular
momentum, while viscous redistribution is weak in quiescence, leading to a small outer region where the orbital angular
velocity \( \Omega\propto r^{-2}\) rather than \(r^{-3/2}\).  It is shown that a short radial scale length ($H$) can
lead to resonances where \(c/H\sim \Omega\).  Such resonances are familiar in asteroseismology but seem to have received
little attention in connection with CV disks.  In \S\ref{sec:sharp_edge} we consider non-resonant excitation at
a discontinuous surface density, which would require a hot atmosphere to maintain continuity of
the pressure.
In \S\ref{sec:nonlinear_effects}, brief attention is given to nonlinear excitation of waves by streamline crossing or by
the impact of the stream from L1, but we argue that these mechanisms should be ineffective in quiescent disks.  We apply
our results to CV disks and circumplanetary disks and link our results to observations in Section \ref{sec:application}.
Section \ref{sec:summary} summarizes our results.

\section{METHODS}\label{sec:waves}
\subsection{Disk models}
Let the the mass of the primary and the binary companion be $M_1$ and $M_2$ respectively and assume that their orbits
are circular.  The orbital frequency is therefore $\omega = \sqrt{G(M_1+M_2)/D^3}$ where $D$ is their separation.  Since
we work with linear theory, the mass ratio \(q\equiv M_2/M_1\) is otherwise treated as infinitesimal: thus the
gravitational potential within the disk is simply \(-GM_1/r\), the axisymmetric part of the companion's tide being
neglected.  We choose units such that $G=M_1=\omega=1$.  The unit length $R_0\equiv (\omega^2/GM_1)^{1/3}$ is therefore
the radius of corotation with the tidal potential.  We use $\Sigma_d$ to denote the typical surface density of the disk;
the exact definition of $\Sigma_d$ differs among models.

The unperturbed equilibrium disk is axisymmetric with outer radius $r_{\rm max}$ and inner radius $r=0$.
In isothermal models, the surface density cannot completely vanish, but there is a characteristic radius $r_0$ beyond
which the surface density \(\Sigma_0\) declines swiftly.
We presume that the structure of the inner parts of the disk is unimportant
for the excitation of waves and for the angular-momentum flux that they carry, as long as the waves
are not appreciably reflected by the inner boundary: short wavelengths subject the ingoing waves to dissipation by
cooling, and nonlinear steepening into shocks.
We assume that $r_{\rm max}<R_0$: i.e., the disk is always truncated inside the corotation resonance. 
This is a realistic assumption for CV disks,  and probably for circumplanetary ones also.  In fact, 
the specific angular momentum of the stream from the L1 point is approximately conserved until it strikes the
disk,\footnote{in fact reduced by 10-20\%, depending on $q$, so that eq.~\eqref{eq:rcirc} is a mild overestimate \citep{Flannery75}}
so the outer radius of the disk is given by the circularization radius
\begin{equation}
  \label{eq:rcirc}
  \frac{r_{\rm c}}{D}\approx (1+q)\left(\frac{r_{\rm L1}}{D}\right)^4.  
\end{equation} Here $r_{\rm L1}$ is the radius of the L1
point, and \( q\equiv M_2/M_1\) the mass ratio.  The inferred radius of the hot spot in quiescent disks
appears to be roughly consistent with eq.~\eqref{eq:rcirc} \citep{Smak71,Stanishev04,BBO16}, indicating at least
partial suppression of the disk viscosity, which would tend to cause the disk to spread outward.  When the disk edge
coincides with the circularization radius~\eqref{eq:rcirc},  it lies well within the ILR
[\(r_{\rm L1}\approx 0.63(1+q)^{-1/3}D\)], let alone the co-rotation radius, unless the mass ratio is extreme
(\(q\lesssim 0.004\)).

The disk profile can be described by the unperturbed surface density $\Sigma_0$ and enthalpy\footnote{Technically, $K_0$
  represents the thermodynamic enthalpy only when the disk is isentropic.}  $K_0 = \int\Sigma_0^{-1}dP_0$.
Force balance relates the latter to the rotation profile:
\begin{equation}
  \label{eq:dK0dr}
\frac{dK_0}{dr} = r\Omega^2 - \frac{GM_1}{r^2}\,.
\end{equation}
When the disk is not isentropic, radial stratification
is measured by the Brunt-V\"ais\"al\"a frequency
\begin{equation}\label{eq:BVdef}
N_0^2 \equiv
  \frac{dK_0}{dr}\left(\frac{d\ln\Sigma_0}{dr} - \frac{1}{c^2}\frac{dK_0}{dr}\right).
\end{equation}
\subsection{Linearisation}
Perturbations can be characterized by the Lagrangian radial
displacement $\xi$ and the Eulerian enthalpy perturbation $K=P/\Sigma_0$ where $P$ is the (vertically integrated) Eulerian pressure perturbation.
For adiabatic perturbation with azimuthal number $m$, the equations governing their pattern in an inviscid disk are
\begin{subequations}\label{eq:eoms}
\begin{align}
\frac{d\xi}{dr} =& -\left[\frac{2m\Omega}{r\sigma} + \frac{1}{r} + \frac{1}{c^2}\frac{dK_0}{dr}\right]\xi +
                   \frac{m^2}{r^2\sigma^2}(K+W_m) - \frac{1}{c^2}K,\label{eq:eom1}\\
\frac{dK}{dr} =& \frac{2m\Omega}{r\sigma}(K+W_m) + \left(\sigma^2-\kappa^2-N_0^2\right)\xi - \frac{dW_m}{dr}\nonumber\\
&-N_0^2\left(\frac{dK_0}{dr}\right)^{-1}K.\label{eq:eom2}
\end{align}
\end{subequations}
The isentropic version of these equations can be found in \citetalias{SPL94}.  Here $\sigma = m(\omega-\Omega)$ is the
frequency of the tide in the corotating frame, and $\kappa$ is the epicyclic frequency, equal to $\Omega$ when
$\Omega\propto r^{-3/2}$.  $W_m$ is the \(m^{\rm th}\) azimuthal harmonic of the tidal potential, for which we take the
$m=2$ quadrupolar approximation
\begin{equation}\label{eq:Wmdef}
W_2 = -\frac{3}{4}\frac{GM_2}{D^3}r^2  
\end{equation}
in all of our numerical calculations.
The \(t,\phi\) dependence \(\exp[im(\phi-\omega t)]\) is implicit for all linearised quantities.
We ignore the self gravity of the disk; the validity of this approximation is discussed
in Appendix \ref{appendix:self_gravity}.
\subsection{The homogeneous problem}
It is convenient to represent the dependent variables by a column vector
\begin{equation}\label{eq:y_def}
\mathbf y \equiv \begin{bmatrix}\sqrt{r\Sigma_0}\xi\\\sqrt{r\Sigma_0}(K+W_m)\end{bmatrix}
\end{equation}
so that the system \eqref{eq:eoms} can be written as \( d\mathbf{y}/dr = \mathbf{My}\) when \(W_m\to 0\).  The factors of
\(\sqrt{r\Sigma_0}\)  ensure that the \(2\times 2\) matrix \(\mathbf{M}\) is traceless and therefore has
eigenvalues \(\pm\sqrt{-k_0^2}\), with
\begin{equation}\label{eq:k}
k_0^2 \approx \frac{\sigma^2-\kappa^2-N_0^2}{c^2}-\left[\frac{1}{2H}+N_0^2\left(\frac{dK_0}{dr}\right)^{-1}\right]^2.
\end{equation}
Here \(H\equiv (-d\ln\Sigma_0/dr)^{-1}\) is the density scale length. 
Exact expressions for  $k_0^2$ and \(\mathbf{M}\) are given in Appendix \ref{appendix:eigenvalues}.
Where $k_0$ varies slowly (\(dk_0/dr\ll k_0^2\)), it becomes the WKBJ wavenumber.
Homogeneous solutions are then oscillatory for \(k_0^2(r)>0\) and evanescent where \(k_0^2(r)<0\).
Radii at which \(k_0^2(r)=0\) are turning points and can be interpreted as resonances (\S\ref{sec:resonances}).
Where \(k_0\) varies rapidly, however, intuition based on WKBJ may fail utterly.
This happens at the edges of our polytropic disks (\S\ref{sec:polytropic_disk}).

Where stratification and density gradients are small,  i.e. well away from the edge, 
\begin{equation}\label{eq:k_approx}
 k_0^2 \approx  \frac{\sigma^2-\kappa^2}{c^2}.
\end{equation}
This form of $k_0^2(r)$ vanishes at the ILR.  Where $k_0^2$ is positive and large, WKBJ is applicable and the radial
group velocity \(V_{\rm g}= -m(\partial k_0/\partial\omega)^{-1}\) is unambiguous.  We choose a basis for the
homogeneous solutions consisting of an outgoing wave \(\mathbf{y}_+\) and ingoing wave \(\mathbf{y}_-\).  Interior to
the ILR, i.e. where \(\omega < \Omega-\kappa/m\), the ingoing wave has a phase that increases with radius,
i.e. \(d\mathbf{y}_\pm/dr\approx \mp ik_0 \mathbf{y}_\pm\) for \(k_0>0\). These solutions can be continued outward to the turning point and
into the evanescent zone \(k_0^2(r)<0\) by solving the homogeneous form of the system \eqref{eq:eoms}.  The phases
and amplitudes of \(\mathbf{y}_\pm\) can be scaled so that they are complex conjugates and so that the solution that
decays outward in the evanescent zone is \(\mathbf y_R \equiv {\rm Real}(\mathbf y_\pm) = (\mathbf y_++\mathbf y_-)/2\).

The angular-momentum flux carried by a wave is
\begin{equation}\label{eq:F_def}  
F = \pi r^2\Sigma_0{\rm Real}\langle v_r^*v_\phi\rangle,
\end{equation}
where $v_r,v_\phi$ are the velocity perturbation given by (for the homogeneous solution)\footnote{When calculating $\langle v_r^*v_\phi\rangle$, we directly use the complex $v_r,v_\phi$ we get from complex $\xi,K$; if we use the real part instead the result should be smaller by a factor of 2.}
\begin{equation}
v_r = -i\sigma\xi, ~~~v_\phi = -\frac{\kappa^2}{2\Omega}\xi+ \frac{m}{r\sigma}K. 
\end{equation}
For a solution to the forced equation ($W_m\neq 0$), $K$ in the above equation is replaced by $(K+W_m)$.  For
homogeneous solutions, $F$ is independent of $r$; the angular-momentum flux is conserved.  We normalize
$\mathbf y_\pm$ so that the angular-momentum flux they carry is
\begin{equation}\label{eq:F0def}
F_0 = \Sigma_dR_0^4\omega^2\left(\frac{M_2}{M_1}\right)^2\left(\frac{D}{R_0}\right)^{-6},
\end{equation}
where $\Sigma_d$ is a constant characterizing the typical density of the disk.
Thus a wave that approaches \(A_{\rm in}\mathbf y_{-}\) as \( r\to 0\) carries a flux \(F_J = |A_{\rm in}|^2 F_0\).

\subsection{The boundary conditions}

For models having a definite edge at \(r_{\max}\), a free outer boundary condition is imposed:
\begin{equation}\label{eq:free_boundary}
\left. K + \xi\frac{dK_0}{dr}\right|_{r_{\max}} =  0\,.
\end{equation}
If the disk has a finite surface density at $r_{\max}$, the free boundary condition requires that the Lagrangian pressure perturbation vanishes there; otherwise, the outer edge is a regular singular point of the linear differential
equations, and the free boundary condition selects the regular solution.
For models having no definite edge, wavelike
part of the solution should decay outward in the evanescent zone.  Both conditions express the physical requirement that
the angular-momentum fluxes emitted by annuli near the boundary tend to zero with their masses.

Near \(r=0\), we require that there is no outgoing wave. This is not always equivalent to saying that the solution
approaches a purely ingoing wave: there is always a locally forced non-wave component of the response to the tide, and this
dominates even at small $r$ if \(\Sigma_0\) increases rapidly inward. (The non-wavelike response is in phase with the
tide and hence carries no flux.) For all models in this paper where it is necessary to examine the behavior as
\(r\to0\) explicitly, however, it suffices to require that \(\mathbf{y}\propto \mathbf{y}_-\) as \(r\to0\).

\subsection{Formal solution for the wave amplitude}\label{subsec:formal_solution}
The first-order system \eqref{eq:eoms} can be recast as a second-order equation for $y_1=\sqrt{r\Sigma_0}\xi$ of the form
\begin{equation}\label{eq:eom_y1}
\frac{d^2}{dr^2}y_1 + p(r)\frac{d^2}{dr^2}y_1 + k^2(r)y_1 = f(r).
\end{equation}
Here $p\approx d\ln c^2/dr$, $k^2\approx k_0^2$, and
\begin{equation}\label{eq:fdef}
f(r) \approx \frac{1}{c^2}\sqrt{r\Sigma_0}\left(\frac{dW_m}{dr}-\frac{2m\Omega}{r\sigma}W_m\right).
\end{equation}
Exact expressions for $p$, $k^2$ and $f$ are given in Appendix \ref{appendix:local_model1}.
A formal solution of this equation is easy to obtain, and we can use it to calculate the amplitude of the tidally excited density wave.

As always, the solution to the inhomogeneous differential equation can be expressed via
integrals over linearly independent homogeneous solutions.
For the latter we take $y_{1,-}$, the complex-valued ingoing wave, and $y_{1,R}$, the
real-valued homogeneous solution that satisfies the outer boundary condition.
Their Wronskian is
\begin{equation}\label{eq:Wronskian}
\mathcal{W}(r)\equiv y_{1,R}\frac{dy_{1,-}}{dr}-\frac{dy_{1,R}}{dr}y_{1,-}\propto \exp\int\limits_r^{r_0}p(r')dr'.
\end{equation}
Note that $\mathcal W(r)\propto 1/c^2$ approximately.
The solution of \eqref{eq:eom_y1} that satisfies our boundary conditions is then\begin{multline}\label{eq:formal0}
  y_1(r) =  y_{1,-}(r)\int\limits_r^{r_{\max}}\frac{y_{1,R}(r') f(r')}{\mathcal{W}(r')}\,dr'\\
  +\  y_{1,R}(r)\int\limits_{0}^{r}\frac{y_{1,-}(r') f(r')}{\mathcal{W}(r')}\,dr'.
\end{multline}
The amplitude of the ingoing wave at the inner boundary can be read off as
\begin{equation}
  \label{eq:Ain}
  A_{\rm in} \equiv \int\limits_{r_{\min}}^{r_{\max}}\frac{y_{1,R}(r) f(r)}{\mathcal{W}(r)}\,dr\,.
\end{equation}
The angular-momentum flux is then
\( F_J= |A_{\rm in}|^2 F_0\), with \(F_0\) given by eq.~\eqref{eq:F0def}.

For a high-Mach-number disk, \(y_{1,R}(r)\) oscillates rapidly, on a scale \(k_0^{-1}\sim c/\Omega\), whereas the rest
of the integrand of eq.~\eqref{eq:Ain} varies slowly. Such integrals tend to be exponentially small (as can be
demonstrated formally by distorting the integration contour into the complex $r$ plane) except for possible
contributions from resonances where \(k_0\to0\) or from radii where the background disk is not smooth. A realistic disk
might have sharp changes even after azimuthal and temporal averaging due to thermal or ionization fronts, standing
shocks at the hot spot, etc. In the idealized models considered here, the only sharp changes occur at the boundary
(\S\ref{sec:polytropic_disk} \& \S\ref{sec:sharp_edge}).
Other models have resonances associated with density or entropy gradients, or to the ILR itself
if that lies within the disk (\S\ref{sec:resonances}). In summary, we expect for all high-Mach-number disks that
significant wave excitation---by which we mean wave fluxes that scale algebraically rather than exponentially with
\(\mathcal{M}\)---should arise locally near such sharp features or resonances.\footnote{The propagation of the waves,
  once excited, may be global.  This depends on nonlinear and dissipative effects that we do not address here.  We
  simply assume that such effects are unimportant within the first wavelength or so of the excitation region.}
The remainder of this paper is dedicated to demonstrating this through examples.

\section{Polytropic disk}\label{sec:polytropic_disk}
As a first example, we consider an isentropic disk with a rotation profile that deviates from keplerian by a
constant factor \(1-\eta\), with \(\eta\ll 1 \):
\begin{equation}
\Omega^2 = \frac{1-\eta}{r^3}.
\end{equation}
Assuming a polytropic equation of state with Emden index $n$ and adiabatic exponent $\gamma=1+1/n$,
\begin{align}\label{eq:polyK}
K_0 &= \eta\left(\frac1r-\frac{1}{r_{\rm max}}\right),\\
\Sigma_0 &= \Sigma_d \left(\frac1r-\frac{1}{r_{\rm max}}\right)^{n},\\
c^2 &= n^{-1}\eta\left(\frac1r-\frac{1}{r_{\rm max}}\right).
\end{align}
Thus the local Mach number \(r\Omega/c\) diverges at the outer edge but 
approaches the constant \(\mathcal{M}_0\equiv\eta^{-1/2}\) at \(r\ll r_{\max}\).

For a given tidal perturbation \( W_m\), the ingoing-wave amplitude $|A_{\rm in}|$ depends on three parameters:
\(\eta\), \(r_{\rm max}\), and $\gamma$. We study the scaling of the result with respect to these, especially the Mach
number \(\mathcal{M}_0=\eta^{-1/2}\).  The only possible resonance in this model is the ILR, and wave excitation depends
strongly on whether or not the ILR occurs inside the disk, as we demonstrate below.

\subsection{ILR inside the disk}\label{subsec:ILR_inside}
When the ILR lies inside the disk, the wave excitation is dominated by the resonance. The angular-momentum flux at the
inner edge of the disk can be analytically estimated (\citealt[hereafter GT79]{GT79}; see also \S\ref{subsec:ILR}):
\begin{equation}\label{eq:GT79_flux}
F_{\rm  in} = 2\pi^2 \left\{\left|\frac{\Sigma_0}{r}
    \left[\frac{d}{dr}(\sigma^2-\kappa^2)\right]^{-1}\right|\left(r\frac{dW_2}{dr}
-\frac{4\Omega}{\sigma}W_2\right)^2\right\}_{\rm ILR}.
\end{equation}
 When the resonance happens at $\Sigma_0\sim\Sigma_d$, this corresponds to a
dimensionless amplitude \(|A_{\rm in}| \sim 1\).   According to eq.~\eqref{eq:GT79_flux}, $F_{\rm in}$ and $|A_{\rm in}|$ do not depend on the
disk Mach number, but only on the local surface density and orbital frequency near the resonance.

When the ILR is well inside the disk, the numerical results in Figures \ref{rmax} and \ref{eta} differ from the analytic
prediction \eqref{eq:GT79_flux} by no more than a few percent.

\subsection{ILR outside the disk}\label{subsec:ILR_outside}

When the ILR lies outside the disk, the excitation should mainly take place near the outer edge, since that is where the
WKBJ wavelength is longest compared to the scale on which the background disk properties vary.  A local analytical
approximation predicts that as the Mach number increases, the wave amplitude decreases faster than any power of
$\mathcal M_0$ (Appendix \ref{appendix:edge}).  We have tested this numerically.

\subsubsection{Numerical results}
\begin{figure}
\centering
\includegraphics[width=\linewidth]{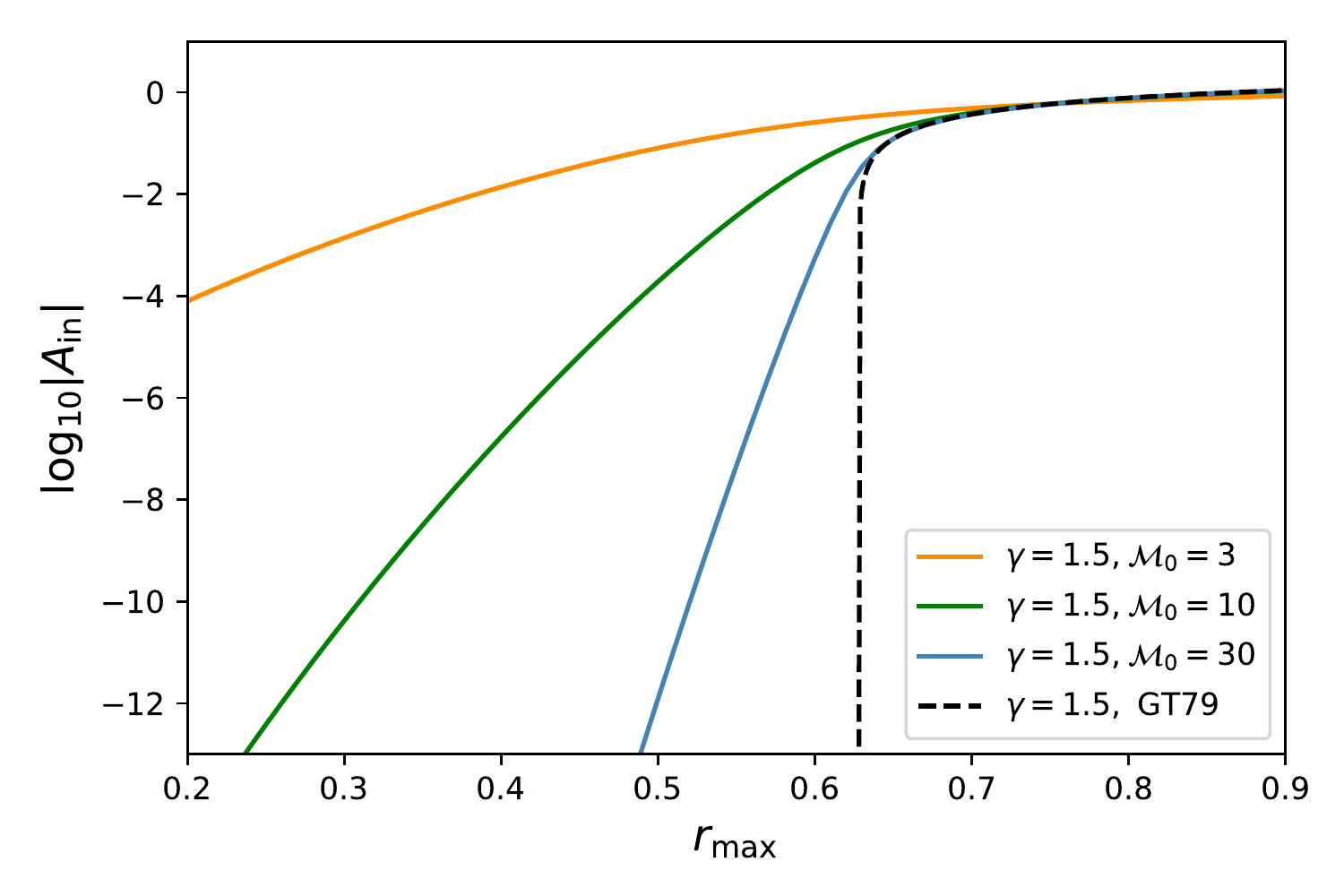}
\caption{Scaling of wave amplitude $|A_{\rm in}|$ versus disk outer radius $r_{\rm max}$
  for the model of \S\ref{sec:polytropic_disk}.
  Dashed curve is the prediction of GT79 [eq.~\eqref{eq:GT79_flux}], which agrees well when
  \(r_{\max}>r_{\textsc{ilr}}\approx 0.63\) so that the ILR lies within the disk. When \(r_{\max}<r_{\textsc{ilr}}\), the amplitude decays
  exponentially with $r_{\rm max}$, and more quickly at higher disk Mach numbers (\(\mathcal{M}_0\)).}
\label{rmax}
\end{figure}

\begin{figure}
\centering
\includegraphics[width=\linewidth]{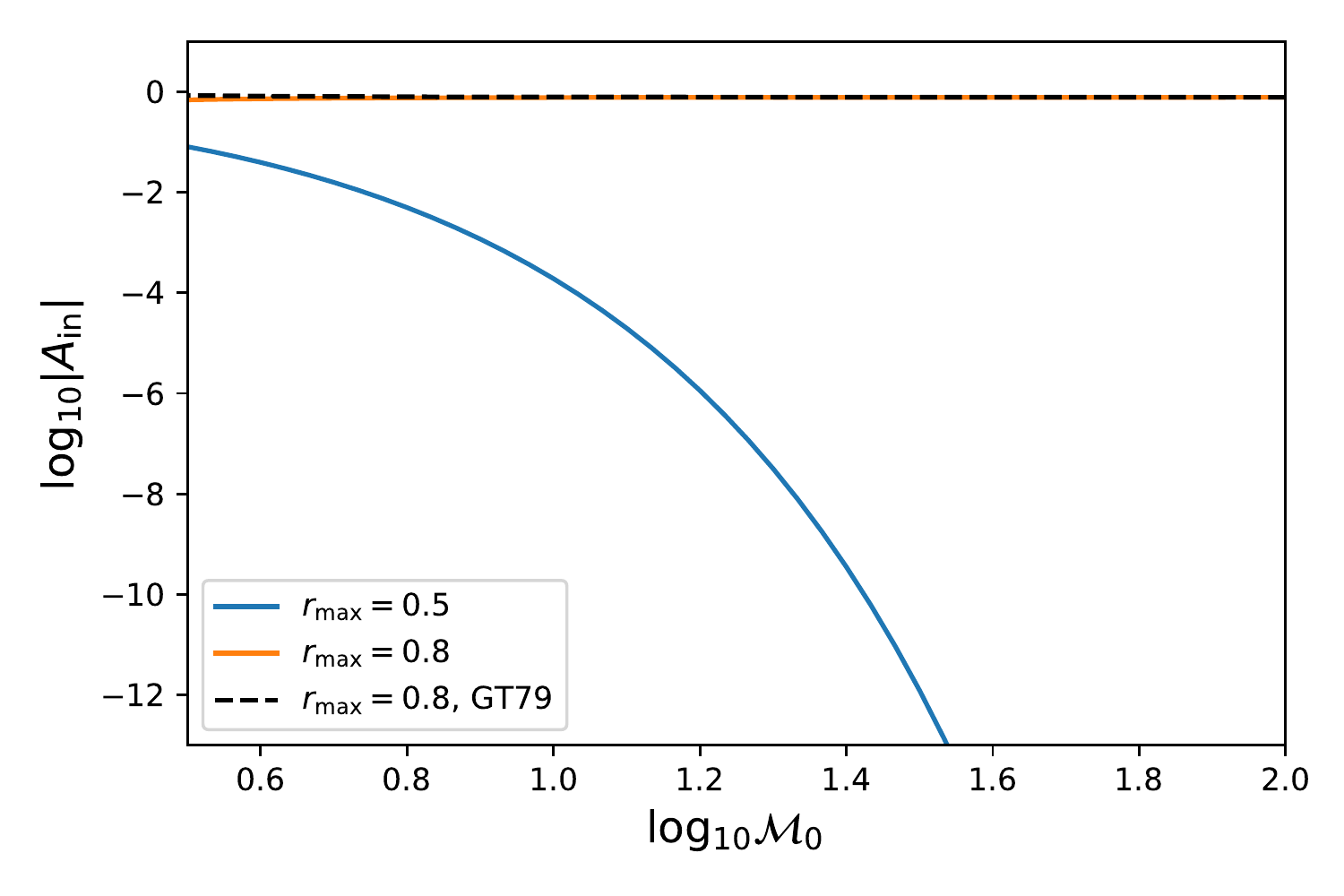}
\caption{Like Fig.~\ref{rmax}, but showing scaling with disk Mach number \(\mathcal{M}_0\).
{\it Blue line:} IRL outside disk, $r_{\rm max}=0.5$.  {\it Orange line:} ILR within disk, $r_{\max}=0.8$. {\it Dashed line}:
GT79 formula [eq.~\eqref{eq:GT79_flux}].}
\label{eta}
\end{figure}

The scaling of $|A_{\rm in}|$ with respect to $r_{\rm max}$ for different $\eta$ and $\gamma$ is shown in Figure
\ref{rmax}.  When the ILR is outside the disk, $|A_{\rm in}|$ decays approximately exponentially as $r_{\rm max}$
decreases.  The rate of this decrease is affected by $\eta$ and $\gamma$; for smaller $\eta$ (i.e. larger Mach number)
and smaller $\gamma$, the amplitude decreases faster as $r_{\rm max}$ decreases.  GT79's result \eqref{eq:GT79_flux}
should be exact as $\eta\to0$ (\(\mathcal{M}_0\to\infty\)).

The scaling of $|A_{\rm in}|$ with respect to characteristic disk Mach number $\mathcal{M}_0\equiv\eta^{-1/2}$, is shown
in Figure \ref{eta}.  When the ILR is outside the disk, the fact that the slope in the log-log plot increases constantly
as $\eta$ decreases suggests that $|A_{\rm in}|$ decays faster than any power of $\eta$, which confirms our analytic
estimates in Appendix~\ref{appendix:edge}.

\subsubsection{An empirical formula for the amplitude scaling}

The scaling of the amplitude when the ILR is outside the disk can be summarized by the following empirical formula,
\eq{
\log_{10}|A_{\rm in}| \approx -0.62\eta^{-1/2}\gamma^{-1}\Delta,\label{A_empirical}
}
where $\Delta$ is a function of $r_{\rm max}$ indicating the deviation of the disk edge from the ILR, and is given by\footnote{$\Delta$ can also be expressed in terms of $r_{\rm max}$ and $r_{\rm ILR}$ (the location of the ILR). Here we choose to express it in terms of frequencies because realistic systems do not always have well-defined $r_{\rm ILR}$ and $r_{\rm max}$.}
\eq{
&\Delta \equiv \left[\frac{\sigma^2-\kappa^2}{\Omega^2}\left(\frac{\Omega}{\omega}\right)^{1/6}\right]_{r_{\rm max}}.\label{eq:Delta_def}
}
The sign of $\Delta$ indicates whether the ILR is inside the disk. This empirical formula is only for the ILR outside the disk, i.e. $\Delta>0$. Note that $\Delta \sim (r_{\rm ILR}-r_{\rm max})$ when the ILR is relatively close to the disk edge.

We find that the empirical formula \eqref{A_empirical} gives a very good fit for the $r_{\rm max}$ dependence
and captures most of
the $\eta$ and $\gamma$ dependence when $|A_{\rm in}|$ is small. 

\subsubsection{The location of wave excitation}

We have numerically investigated the radius at which most of the excitation occurs (\(r_{\rm eff}\)) by
adding a small imaginary part to $\omega$, which corresponds physically to a tidal perturbation that is growing with
time. 
The wave amplitude near the inner edge is then reduced in proportion to the group delay from the point of excitation.
By these means, we find that $r_{\rm max}-r_{\rm eff} \lesssim \eta$.

\subsection{Summary of this section}
When the ILR is inside the disk, GT79's formula \eqref{eq:GT79_flux}
accurately predicts the angular-momentum flux, independently of the disk Mach number and other disk
properties absent from that formula.  When the ILR is outside the disk, the excitation happens near the outer edge
and is exponentially weak, as codified by the empirical formula \eqref{A_empirical}.  We expect similar
behavior from all models where the temperature falls linearly to zero at the disk edge,
and no resonance or discontinuity exists within the disk.

\section{Wave excitation by resonances}\label{sec:resonances}
Even in two-dimensional disks, other resonances besides the ILR can exist, and in fact should exist near the disk edge
provided that the radial scale lengths of density and pressure vary slowly there.  These resonances tend to be weaker
than the ILR.  Although familiar in asteroseismology (e.g., the acoustic cutoff for p-modes), they seem not to have been
studied before in this context.

We will start by developing a local model for resonances and obtain a general analytical formula for the angular
momentum flux they produce in the limit of high Mach number (short wavelengths).  We apply the results to different
types of resonances, and discuss the scaling of the angular-momentum flux with Mach number for each case.
\subsection{A local model for resonances}
We identify resonant radii with zeros of the determinant $k_0^2$ of the matrix $\mathbf M$ in the first-order linear
differential system \eqref{eq:y_first_order}.  The local analysis of wave excitation at resonances offered below is
justified only if relevant properties of the disk vary slowly near the resonance.  The
polytropic disks of the previous section formally have an acoustic-cutoff resonance (\S\ref{subsec:ACR}), but only very
close to the disk edge ($r_{\max}$), where $H=(-d\ln\Sigma_0/dr)^{-1}\approx (r_{\max}-r)/\eta$ varies quickly;
evidently, the analysis below does not apply to such cases.

Consider the second order equation \eqref{eq:eom_y1} for $y_1=\sqrt{r\Sigma_0}\xi$ near a zero of $k_0^2$.\footnote{Here
  we assume that the zero of $k^2$ is so close to the zero of $k_0^2$ that they do not have to be distinguished from
  each other.}  Let $r_{\rm res}$ be the location of this zero.  If $k_0^2$ varies smoothly, the following length scale
can be associated with the resonance:
\begin{equation}\label{eq:lambda}
\lambda\equiv\left|\frac{dk_0^2}{dr}\right|^{-1/3}_{r_{\rm res}}
\end{equation}
When this is small compared to the radius and to the scales of variation of relevant disk properties---including
$H$ but not necessarily $\Sigma_0$ itself---we can approximate \eqref{eq:eom_y1} by the simplified local
model
\begin{equation}\label{eq:airy}  
\frac{d^2y_1}{dx^2} - xy_1 = \lambda^2f(r_{\rm res}) e^{-\alpha x}
\end{equation}
Here $x \equiv (r-r_{\rm res})/\lambda$ and
\begin{equation}
  \label{eq:alphadef}
\alpha\equiv\frac{\lambda}{2H}\,.
\end{equation}
The derivation of eq.~\eqref{eq:airy} is given in Appendix \ref{appendix:local_model2}.
The exponential term is usually not needed to describe excitation at an ILR, where $\Sigma_0$ varies slowly,
but it is needed for the resonances of \S\ref{subsec:ACR} \&\ref{subsec:SACR}, in which the density scale length may be
comparable to $\lambda$, so that $\alpha\sim \mathcal O(1)$.

The solution for the amplitude and flux of the ingoing wave is given by GT79 for $\alpha=0$, and
can be obtained by the methods of \S\ref{subsec:formal_solution} with the appropriate replacements.
This treatment would extend to $\alpha<0$ because the integral analogous to eq.~\eqref{eq:Ain} for the
amplitude of the ingoing wave would then be proportional to
\begin{equation}\label{eq:Ai-exp-int}
  \int\limits_{-\infty}^{\infty} \mathrm{Ai}(x)e^{-\alpha x}\,dx = e^{-\alpha^3/3}\qquad \mbox{Real}(\alpha) <0.
\end{equation}
Here $\mathrm{Ai}(x)$ is the Airy function that decays \(\sim\exp(-\tfrac{2}{3}x^{3/2})\) as $x\to+\infty$ and is
oscillatory for $x<0$.
We need $\alpha\ge0$ since $d\ln\Sigma_0/dr<0$, however.  The integral above is then not convergent, so we
solve eq.~\eqref{eq:airy} by other means in Appendix \ref{appendix:local_model2}.
It turns out that the amplitude $A_{\rm in}$ of the ingoing wave is exactly as if
eq.~\eqref{eq:Ai-exp-int} held for positive as well as negative $\alpha$.
The angular-momentum flux is therefore
\begin{align}\label{eq:Fin_res_general}
F_{\rm in} &\approx m\pi^2 \frac{r\Sigma_0}{c^2} \left|\frac{dk_0^2}{dr}\right|^{-1} 
|W'_m|^2 e^{-2\alpha^3/3}\,,\nonumber\\
\mbox{where}\quad W'_m&\equiv \frac{dW_m}{dr}-\frac{2m\Omega}{r\sigma}W_m\,,
\end{align}
all quantities being evaluated at the resonance.
Note that the frequent combination hereby abbreviated as $W'_m$ is not
simply the radial derivative of the tidal potential $W_m$.

Depending on which terms in $dk_0^2/dr$ are dominant, the scaling of $F_{\rm in}$ can be different. Below, we will
consider the scaling of $F_{\rm in}$ for different resonances.

\subsection{Lindblad resonance}\label{subsec:ILR}
When stratification and density gradient are both negligible, the resonance is located approximately at
$\sigma^2-\kappa^2=0$.
In this case,
\begin{equation}
\frac{dk_0^2}{dr}\approx \frac{1}{c^2}\frac{d(\sigma^2-\kappa^2)}{dr}\,,
\end{equation}
and the angular-momentum flux is
\begin{equation}\label{eq:res_flux}
F_{\rm in} \approx m\pi^2 r\Sigma_0\left|\frac{d}{dr}(\sigma^2-\kappa^2)\right|^{-1}|W'_m|^2\,.
\end{equation}
This is identical to GT79's result \eqref{eq:GT79_flux}, although they also included
disk self gravity.  The factor \(\exp(-2\alpha^3/3)\) has been omitted on the presumption that $H\gg\lambda$.
\subsection{Acoustic-cutoff resonance (ACR)}\label{subsec:ACR}
In an isentropic disk, or one with negligible radial stratification, \eqref{eq:k} at high Mach number becomes
\begin{equation}\label{eq:k0_ACR}
k_0^2\approx \frac{\sigma^2-\kappa^2}{c^2} - \frac{1}{4H^2}.
\end{equation}
The resonance occurs where this vanishes,
i.e. where the acoustic cutoff frequency of the disk $\omega_{\rm ac} = c/2H$ 
matches $\sqrt{\sigma^2-\kappa^2}$. Therefore we refer to this
resonance as an acoustic-cutoff resonance (ACR).  This requires $H\sim c/\Omega$ (the vertical scale height) unless
$\sigma^2-\kappa^2\ll\Omega^2$, in which case the resonance reduces essentially to an ILR.  The natural
location for an ACR proper is in the outermost parts of the disk where $\Sigma_0$ declines rapidly.

\subsubsection{Scaling of angular-momentum flux with $\mathcal{M}$}\label{subsubsec:M_scaling_ACR}
The angular-momentum flux has the form \eqref{eq:Fin_res_general} but with $k_0^2$ as in eq.~\eqref{eq:k0_ACR} rather than
eq.~\eqref{eq:k_approx}.

Consider a fixed subkeplerian rotation curve $\Omega(r)<\Omega_K(r)$.  For an isothermal equation of state, since the
pressure gradient must make up the difference between centrifugal force and gravity,
\(H=r(\Omega_K^2-\Omega^2)/c^2\propto\mathcal{M}^{-2}\) at a fixed radius.  The ACR will occur where the two terms on
the right side of eq.~\eqref{eq:k0_ACR} balance, whence \(H_{\textsc{acr}}\propto\mathcal{M}^{-1}\) rather than
\(\mathcal{M}^{-2}\). Thus we need $(\Omega_K-\Omega)_{\rm ACR}\sim \mathcal M^{-1}$.

If the density scale length varies slowly, i.e. $0<-dH/dr\ll 1$, then the flux excited at an ACR will tend to decrease rapidly
with increasing Mach number.  Let \(r_0<r_{\textsc{acr}}\) be some radius interior to the ACR where the
surface density is characteristic of the disk as a whole, but close enough so that the scale lengths at the two radii
are comparable.  Then by the intermediate-value theorem, there exists an intermediate radius
\(\bar r\in(r_0,r_{\textsc{acr}})\) such that
\begin{equation*}
  \ln\left[\frac{\Sigma_0(r_0)}{\Sigma_0(r_{\textsc{acr}})}\right] = \int\limits_{r_0}^{r_{\textsc{acr}}} \frac{dr}{H(r)} = 
\frac{r_{\textsc{acr}}-r_0}{H(\bar r)}
\end{equation*}
If $H$ varies slowly, and since $r/H=\mathcal O(\mathcal{M})$ at the ACR,
the numerator of this last expression can be much larger than the denominator,
whence \(\Sigma_0(r_{\textsc{acr}})\), to which the flux is directly proportional, is small.
Furthermore, if the gradient of $k_0^2$ near the resonance is dominated by the variation of the scale length,
then it follows from eqs.~\eqref{eq:k0_ACR}, \eqref{eq:lambda}, and \eqref{eq:alphadef} that
\(\alpha^3 = \lambda^3/8H^3 \approx (-4 dH/dr)^{-1}\)\,,
and so the factor \(e^{-2\alpha^3/3}\) in the flux \eqref{eq:Fin_res_general} is also small.

\subsubsection{Numerical confirmation}\label{subsubsec:ACR_numerical}
Consider an isothermal disk with rotation profile
\begin{equation}
\Omega^2(r) = \frac{GM_1}{r^4}\left(r^{-\beta} + r_0^{-\beta}\right)^{-1/\beta},
\end{equation}
where $r_0$ and $\beta>1$ are constants. This corresponds to a disk with a keplerian rotation profile for $r\ll r_0$, and
constant specific angular momentum for $r\gg r_0$.  Larger $\beta$ makes for a sharper transition between the two regimes.
To maintain this rotation profile, the density scale height satisfies
\begin{align}
\frac{1}{H} =
  \frac{1}{c^2}\left(\frac{GM_1}{r^2}-r\Omega^2\right) \sim \left\{\begin{array}{ll}0 &(r\ll r_0)\\\mathcal M^2/r
      &(r\gg r_0)\end{array}\right.,
\end{align}
The ACR, which requires $1/H\sim \mathcal M/r$, occurs at $r\sim r_0$.

\begin{figure}
\includegraphics[width=\linewidth]{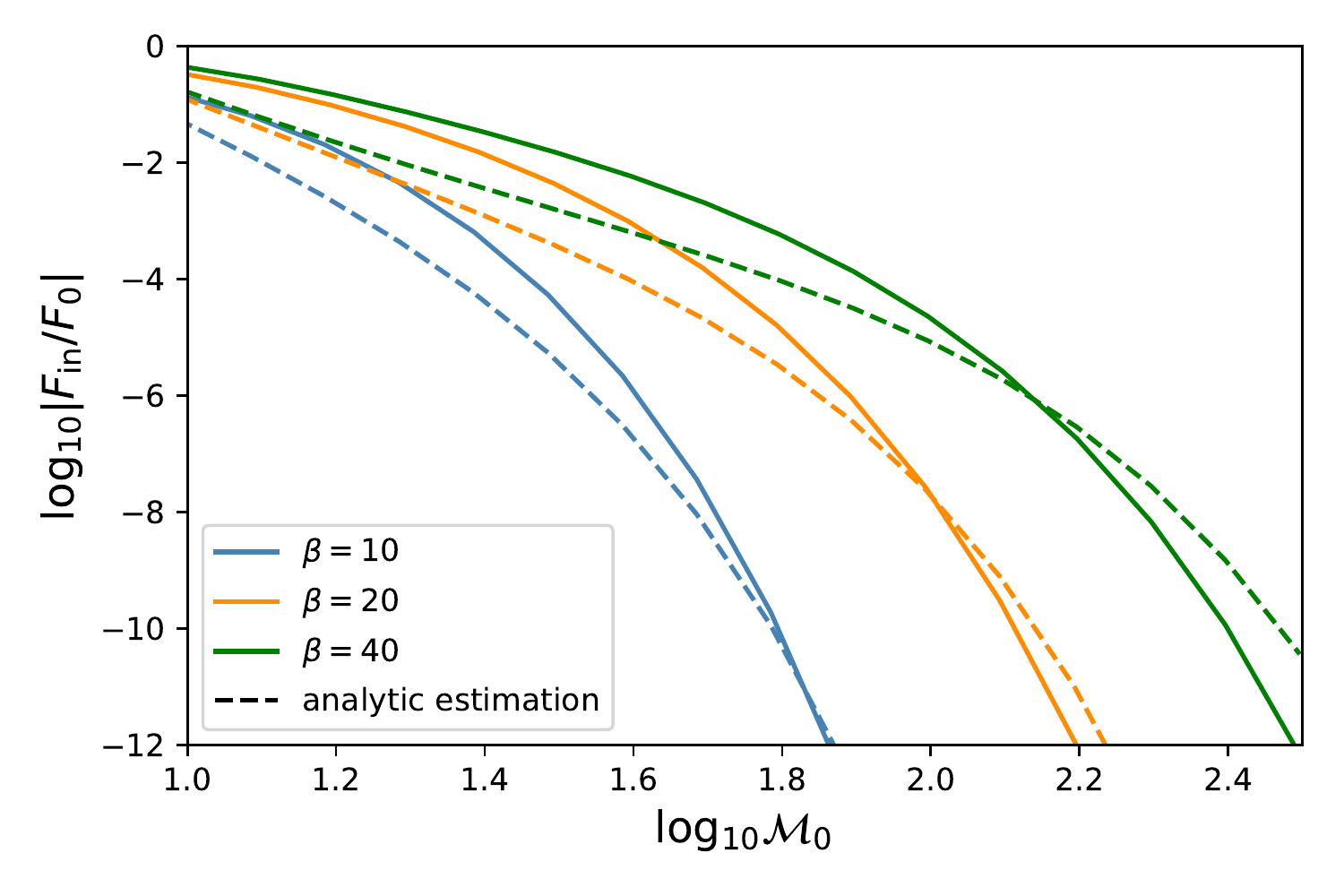}
\caption{The angular-momentum flux $F_{\rm in}$ due to ACR as a function of the characteristic Mach number $\mathcal
  M_0$ for $r_0=0.5$ and different $\beta$ values. The disk model is described in \S\ref{subsubsec:ACR_numerical}, and
the reference flux $F_0$ [eq.~\eqref{eq:F0def}] is calculated for $\Sigma_d \equiv \Sigma_0(0)$.}
\label{fig:beta_model}
\end{figure}

As Figure~\ref{fig:beta_model} shows, the numerical results broadly agree with the analytical prediction
\eqref{eq:Fin_res_general}.  At large Mach number, $\Sigma_0(r_{\rm ACR})$ and the angular-momentum flux fall off
rapidly---faster than any power law; but the curvature in these log-log plots is less for larger $\beta$, i.e. for more
abrupt transitions between keplerian rotation and constant specific angular momentum.

\subsection{The effect of stratification: stratification-shifted ACR (SACR)}\label{subsec:SACR}
When the effect of stratification becomes nontrivial (i.e. when the terms with $N_0^2$ are among the leading order terms in $k_0^2$ near the resonance), the location of the resonance can be shifted.
In this case, we refer to this resonance as a stratification-shifted ACR (SACR).

Since SACR depends on too many parameters ($H,N_0^2,dK_0/dr$ etc.), $F_{\rm in}$ cannot be simplified unless certain
restrictions are imposed on the disk profile.  Let us make the following assumptions:
\begin{itemize}
\item $\gamma\equiv (\partial\ln P/\partial\ln \Sigma)_{\rm ad}$ is $\mathcal O(1)$ and slowly varying.
\item The disk temperature $T\propto P_0/\Sigma_0$ is finite and slowly varying.  With the previous
  condition, the same goes for $c^2$.
\end{itemize}

Under these assumptions, when the ILR lies far outside the disk and $H\ll r$ (which, as we will show later, is required
at the resonance),
\begin{equation}
N_0^2 \sim \frac{c^2}{H^2},~~~\left|\frac{dK_0}{dr}\right|\sim \frac{c^2}{H}.
\end{equation}
As a result, the SACR corresponds to
\begin{equation}
\sigma^2-\kappa^2 \sim N_0^2 \sim \frac{c^2}{H^2} \sim c^2 \left[N_0^2\left(\dfrac{dK_0}{dr}\right)^{-1}\right]^2.
\end{equation}
Thus the $H$ required for SACR is similar to that required for ACR.

If $H$ is one of the fastest varying parameters, then
\begin{equation}
F_{\rm in}\sim \left[\frac{r\Sigma_0}{\sigma^2-\kappa^2}\left|\frac{d\ln H}{dr}\right|^{-1}|W'_m|^2\right]_{\rm SACR}.
\end{equation}
In general, one can replace $d\ln H/dr$ by $d\ln X/dr$ where $X$ is the fastest varying parameter. 
Therefore, the scaling of the flux with $\mathcal{M}$ should be much the same as for the ACR
(\S\ref{subsubsec:M_scaling_ACR}).

\section{Wave excitation at discontinuities}\label{sec:sharp_edge}
In this section, we consider another possible mechanism for wave excitation: discontinuities in the equilibrium disk.
As an example, we consider a discontinuity in the surface density itself---a disk with finite $\Sigma_0(r_{\max})$.  
The rotation profile is keplerian even at the edge, and $k_0r\gg1$, so that WKBJ is applicable everywhere.
Pressure balance in equilibrium would require a tenuous hot atmosphere to confine the disk edge.  This is
probably not realistic, but it seems to be what SPL assumed for their analytic estimates.  It may also serve as
a proxy for sharp changes in surface density within the disk interior, perhaps due to thermal/ionization fronts.
The wave excitation is estimated analytically via a local model, and then confirmed numerically.

\subsection{Discontinuous surface density}\label{subsec:discon_sigma}

The distinguishing feature of this model is that the surface density is nonzero at the edge, as is the sound speed.  So
to study wave excitation, we adopt a local model in which these are constants, whence $dK_0/dr=0$.  For
consistency, the rotation curve exactly balances gravity and is exactly keplerian, but we continue to write
$\kappa^2$ not $\Omega^2$ in eq.~\eqref{eq:k_approx} for $k_0^2$ as a reminder of the ILR.  We
treat $k_0$ as locally constant, assuming that the ILR lies a distance $\gg k_0^{-1}$ beyond $r_{\max}$.  Defining
\(x\equiv (r-r_{\max})k_0\) and \(f_0\equiv f(r_{\max})\), our local
approximation to \eqref{eq:eom_y1} becomes
\begin{equation}
  \label{eq:edge_model}
  \frac{d^2y_1}{dx^2} + y_1 = k_0^{-2}f_0\,.
\end{equation}
The Lagrangian pressure perturbations should vanish if the edge is confined by a tenuous atmosphere of constant
pressure, so we impose the free boundary condition \eqref{eq:free_boundary}.  Since $dK_0/dr=0$, this reduces to
$K(r_{\max})=0$; cast in terms of \(y_1=\sqrt{r\Sigma_0}\xi\), the free boundary condition becomes
\begin{equation}
  \label{eq:edge_bc}
 x=0:\quad \frac{dy_1}{dx} + \frac{1}{k_0r}\left(\frac{2m\Omega}{\sigma}+\frac{1}{2}\right)y_1 =
  \frac{m^2\sqrt{r\Sigma_0}}{k_0r^2\sigma^2}W_m\,.
\end{equation}
The non-wavelike particular solution to eq.~\eqref{eq:edge_model} is simply \(y_{1,\rm nw}= k_0^{-2} f_0\).  Since
\(k_0\propto c^{-1}\) and \(f_0\propto c^{-2}\), this solution is independent of $\mathcal M$, but it does not satisfy the
boundary condition \eqref{eq:edge_bc}.  So we must add to it an ingoing wave \(A_{\rm in}\exp(ix)\) that satisfies the
homogeneous form of eq.~\eqref{eq:edge_model}.  Inserting the sum of these two into the boundary condition yields,
to leading order in $(k_0r)^{-1}$,
\begin{equation}
  \label{eq:Ain_edge}
  A_{\rm in} \approx \left.k_0^{-1}\sqrt{r\Sigma_0}\,\widetilde W_m\right|_{r_{\max}}\,,
\end{equation}
in which the dimensionless quantity
\begin{equation}\label{eq:Wtilde}
  \widetilde W_m\approx \frac{W_m}{r^2\sigma^2}\left[m^2- \frac{\sigma^2+4m\Omega\sigma}{2(\sigma^2-\kappa^2)}
\left(\frac{d\ln W_m}{d\ln r}-\frac{2m\Omega}{\sigma}\right)\right]
\end{equation}
involves the tidal potential but not $c$ or $\Sigma_0$.

\subsubsection{Mach-number scaling}

The angular-momentum flux \(\pi m c^2k_0|y_1|^2\) carried by the ingoing wave in this model is 
\begin{equation}
  \label{eq:sharp_edge_Fin}
  F_{\rm in} = \left.\pi m|\widetilde W_m|^2\frac{c^3 r\Sigma_0}{\sqrt{\sigma^2-\kappa^2}}\right|_{r_{\max}}\,,
\end{equation}
which clearly scales as \(\mathcal{M}^{-3}\). The scaling with $r_{\max}$ at fixed \(\mathcal{M}=(r\Omega/c)_{r_{\max}}\) is
\(F_{\rm in}\propto r_{\max}^{2m+3}\Sigma_0(r_{\max})\) if \(W_m\propto r^m\) and \(r_{\max}\ll r_{\textsc{ilr}}\).

SPL seem to have used a model similar to this one for their analytical estimates, though they are not explicit about the
equilibrium conditions at the disk edge.  They discuss both free and rigid (vanishing radial displacement) boundary
conditions but do not say which they used for their estimates of the angular-momentum flux.  It is easy to see from the
analysis above that the two boundary conditions lead to different scalings with Mach number: For suppose that $y_1=0$ at
the edge.  Then since the non-wave particular solution does not vanish,
\(y_{1,\rm nw}= k_0^{-2} f_0\propto\mathcal{M}^0\), an equal and opposite wavelike part would have to be added to it,
leading to \(F_{\rm in}\propto\mathcal{M}^{-1}\) rather than \(F_{\rm in}\propto\mathcal{M}^{-3}\).

\subsubsection{Numerical confirmation}
\begin{figure}
\centering
\includegraphics[width=\linewidth]{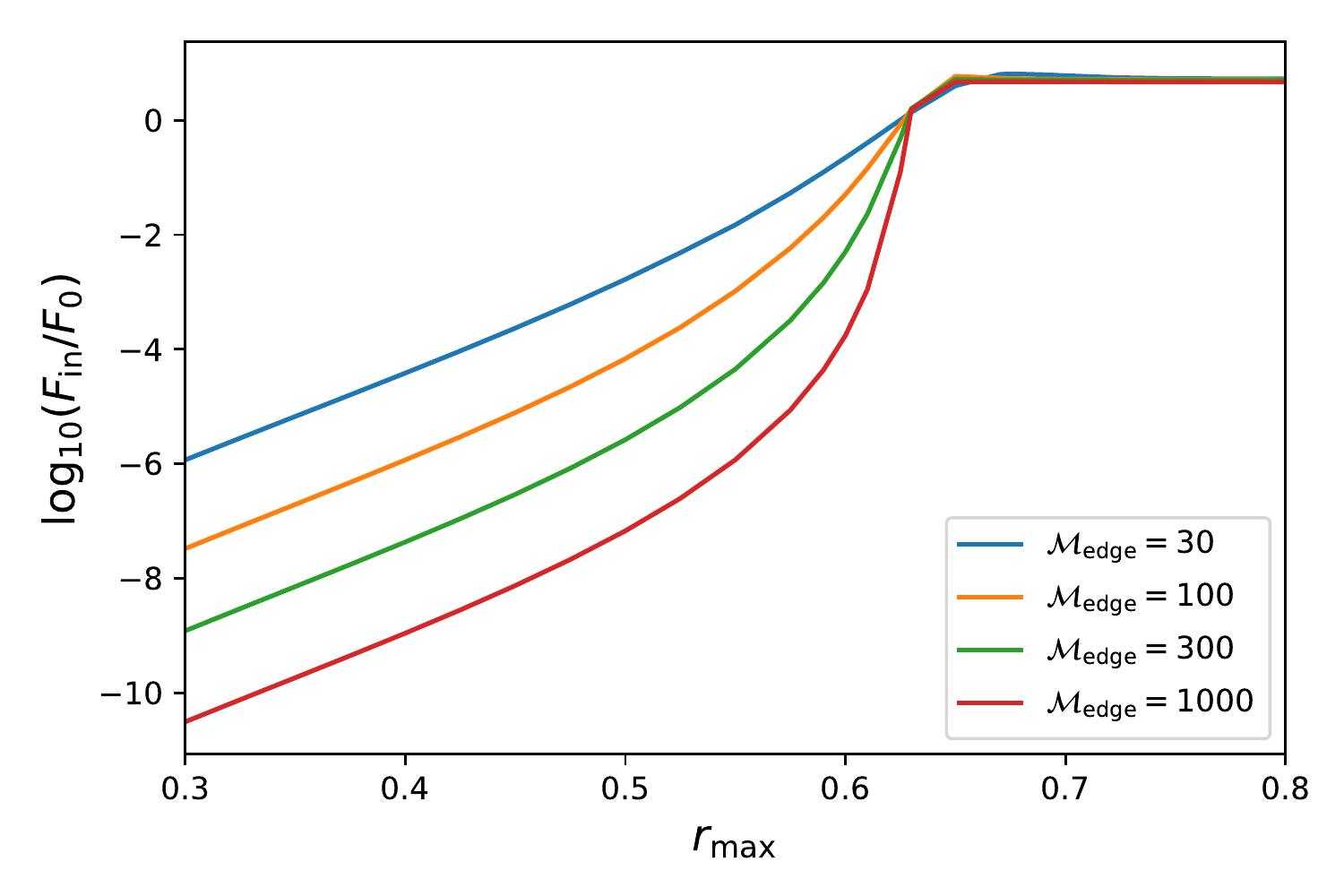}
\caption{Numerical results for the angular-momentum flux in pressure-confined
 disks with uniform density and sound speed.  All models have \(\Sigma_0=\Sigma_d\) but different $c$.
The scaling is roughly \(\sim\mathcal{M}^{-3}\) when the
 ILR lies well outside the disk [eq.~\eqref{eq:sharp_edge_Fin}] but independent of $\mathcal{M}$ when the ILR lies
 inside.}
\label{fig:uniform_disk}
\end{figure}
To test the scaling relations predicted above, we have made numerical calculations for a disk with uniform $c$ and
$\Sigma_0=\Sigma_d$.  These results confirm that when the ILR lies well outside the disk,
$F_{\rm in}\propto \mathcal M^{-3}$, and that when the ILR lies inside the disk, $F_{\rm in}$ is independent of
$\mathcal M$ (Fig.~\ref{fig:uniform_disk}).

Note that there is an intermediate regime at $r_{\rm max}\sim r_{\rm ILR}$ which connects the two limiting cases (ILR
deep inside / far outside the disk). At large $\mathcal M$, $F_{\rm in}$ has a steep dependence on $r_{\rm max}$ in
this intermediate regime.

\section{Nonlinear effects}\label{sec:nonlinear_effects}
In this section, we discuss several nonlinear effects which are relevant to the problem. These nonlinear effects include the truncation of the disk by the companion which tends to keep the disk edge inside the ILR and the mass inflow from the companion which causes angular momentum transport by itself.
\subsection{Disk truncation by the binary companion}\label{subsec:disk_trunction}
In our discussion above we have effectively assumed that the companion mass ratio $q\equiv M_2/M_1$ is infinitesimal. This means that the ILR can be inside the disk. However, this is not the case in the systems we are interested in. For CV disks, $q$ is typically of order $0.1$ to 1, while for circumplanetary disks, $q$ is $\gtrsim 10^3$.
Having a finite $q$ affects the maximum possible size of the disk; especially, the edge of the largest possible disk may still be far from the ILR, in which case the excitation is bound to be weak.

We investigate the maximum possible size of the disk using the method in \cite{Paczynski77}.
The edge of the largest possible disk is defined as the maximum stable periodic streamline around the primary when the pressure and gravity of
the disk is neglected.  We survey a large range of $q$ from $10^{-2}$ to $10^4$, and $\sigma^2-\kappa^2$ at the disk
edge is always positive and comparable to $\omega^2$ (the binary's orbital frequency).\footnote{Here we assume $\kappa=\Omega$ for simplicity. Using a more realistic assumption, that $\kappa^2 = 2\Omega R^{-1}d(R^2\Omega)/dR$ where $2R$ is the diameter of a periodic streamline and $2\pi/\Omega$ the period of it, gives the same qualitative result.}

The size of the disk is also limited by the angular momentum of the accretion stream. In this case, the circularization radius $r_c$ \eqref{eq:rcirc} gives an estimate of the size of the disk. $r_c$ is usually smaller than the maximum possible size of the disk defined above.

One caveat is that we only consider an infinitely thin disk. Disk pressure, self gravity and viscosity may help increase the stability of the disk and make the disk edge closer to the ILR.

\subsection{Shocks launched by the accretion stream}

The impact on the disk edge of the stream of material from the L1 point can also launch shocks.  In a time-averaged
sense at least, these shocks (like the tidally excited ones) have angular pattern speed $\omega$, the orbital angular
velocity of the binary; since this is less than the orbital angular velocity of the disk, 
these shocks also deposit negative angular momentum in the disk.  Here we make a rough estimate of the torque on the disk
due to these shocks.

Suppose that that the outer edge is located at the circularization radius \eqref{eq:rcirc}.
The relative velocity $v_{\rm rel}$ between the stream and the outer edge of the disk where they collide is then
approximately radial, since the stream and the disk edge have similar specific angular momentum.
The stream delivers radial momentum to the disk edge at rate $\approx \dot M_{\rm d}v_{\rm rel}$ where $\dot M_{\rm d}$
is the mass flow in the stream---the rate of addition of mass to the disk.
The impact drives a shock into the disk.  If the shock is weak, it propagates radially approximately at the local sound
speed, $c$.
If the shock carries all or most of the radial momentum of the stream, then its mechanical power is approximately
\(\dot M_{\rm d}v_{\rm rel} c\) in a frame corotating with the disk edge.  The associated torque is found by dividing
the power by the difference in angular velocity between the shock pattern ($\omega$) and the disk edge
[\(\approx\Omega(r_{\rm c})\equiv\Omega_{\rm c}\)], viz.
\(\dot J_{\rm w}\approx \dot M_{\rm d}v_{\rm rel} c (\omega-\Omega_{\rm c})^{-1}\).
As already noted, the torque is negative since \(\omega<\Omega_{\rm c}\).
On the other hand, material angular momentum is added to the disk by the stream at the rate
\( \dot J_{\rm d} \approx \dot M_{\rm d}\Omega_{\rm c}r_{\rm c}^2\).  Hence
\begin{equation}
  \label{eq:Jdot-ratio}
-\frac{\dot J_{\rm w}}{\dot J_{\rm d}} \approx \frac{v_{\rm rel}c}{(\Omega_{\rm
    c}-\omega)\Omega_{\rm c}r_{\rm c}^2} \lesssim(\sqrt{2}-1)\frac{c}{\Omega_{\rm c}r_{\rm c}}\,.
\end{equation}
The last inequality presumes that the total velocity of the stream is approximately the local escape velocity
\(\sqrt{2}\Omega_{\rm c}r_{\rm c}\).  At high Mach number ($c\ll\Omega_{\rm c} r_{\rm c}$), the torque exerted by the
impact of the stream is removes only a small fraction of the angular momentum that the stream itself adds to the disk.

Apart from its dependence on Mach number, the torque exerted by the stream scales with the rate of addition of
mass, whereas the torque exerted by the tide scales with the mass already present in the disk.  Therefore, if the outer
disk were to become sufficiently massive, an equilibrium could in principle be reached between the rate of addition of
angular momentum by the stream and the rate of its removal by the tide (though the required mass may be unreasonably
large if the disk has high Mach number and ends at the circularization radius, rather than the streamline-crossing or
ILR radius).  The inequality~\eqref{eq:Jdot-ratio}  argues that the shocks excited by the stream itself cannot
achieve such an equilibrium at any disk mass.

\section{Applications}\label{sec:application}
\subsection{CV disks in quiescence}
For CV disks in quiescence, observational results \citep{Menou00,CSWSH12} suggest that tidal perturbations may be important for the accretion to the primary and the evolution of the disk.
Here we discuss whether the claim that accretion during quiescence is mainly due to the tidal excitation of density waves is consistent with our theory.

Consider the effective $\alpha$ of the disk, which in the context of a forced and accreting disk can be defined as
\eq{
\alpha = \frac{\dot M}{3\pi\Sigma c h}
}
where $h = c/\Omega$ is the disk scale height.
Assuming that accretion conserves the mean specific angular momentum of the disk, this effective $\alpha$ is related to our dimensionless amplitude $|A_{\rm in}|=(F_{\rm in}/F_0)^{1/2}$ by (for mass ratio $q\lesssim 1$)
\eq{
\alpha \sim q^2|A_{\rm in}|^2\mathcal M^{-2},
}
where $\mathcal M$ is the typical Mach number of the disk. For CV disks in quiescence, observations of disk temperatures suggest a $\mathcal M\sim 100$ to $200$ at quiescence (see, e.g. \citealt{RWPP16}), and the typical mass ratio is $q\sim 0.1$.
The effective $\alpha$ during outburst and quiescence are $\alpha_{\rm hot} \sim 0.1$ \citep{Smak84} and $\alpha_{\rm
  cold}\sim 10^{-3}$ \citep{CSWSH12}, although with considerable variation from one system to another.

In this paper we have discussed three main cases of wave excitation when the ILR does not lie inside the disk:
\begin{itemize}
\item When the disk contains no resonance or discontinuity, the amplitude scales exponentially with respect to the Mach number. For quiescent CV disks, the Mach number is large and the excitation trivial.
\item Resonances other than the ILR (i.e. ACR and SACR) may exist in the disk.
It is possible for them to produce sufficient angular-momentum flux to be the main accretion mechanism. However, in realistic disks, $\Sigma_0$ at the resonance tends to be much smaller than the typical surface density of the disk, reducing the wave excitation.
Therefore, it remains uncertain whether resonant excitation of density waves can produce enough accretion.
\item The disk profile may contain discontinuities. In the context of quiescent CV disks, the accretion stream and inefficient angular momentum transport may cause the disk to have a sharp edge. If the disk edge is sharp enough (i.e. the density drops from some large value to trivially small within a length scale $\ll r/\mathcal M$), this mechanism is very promising. For instance, for a disk with uniform density and sound speed, with $r_{\rm max}=0.5$ the effective $\alpha$ is given by
\eq{
\alpha \approx 0.0015 \left(\frac{\mathcal M(r_{\rm max})}{100}\right)^{-1}.
}
This is comparable to the observed $\alpha_{\rm cold}$.
\end{itemize}
The above discussion suggests that the accretion driven by tidally excited density waves can indeed produce the observed $\alpha_{\rm cold}$ when the disk contains a resonance (ACR or SACR) that happens at a sufficiently large density, or when the disk has a sharp edge.

Our theory is also consistent with the observation that $\alpha_{\rm cold}$ has a steep dependence on the disk size
\citep{CSWSH12}.  At high Mach number, the excitation at an ACR or SACR is much weaker than at an ILR.  Yet the net
excitation is a continuous function of disk radius.  Necessarily therefore, the wave torque varies rapidly with
$r_{\max}$ if this is not much less than $r_{\textsc{ilr}}$.  This effect is illustrated in Fig.~\ref{fig:uniform_disk} for
discontinuous disks but occurs more generally.

Overall, our result is in broad agreement with observation, and it is possible that the accretion during quiescence is mainly driven by tidally excited density waves.
However, since the amplitude can vary by orders of magnitude across different disk models, whether tidally excited density waves are the main cause of accretion remains an open question until the disk profile and evolution can be determined with high accuracy.
Our results also suggest that extra caution has to be taken when modelling the disk in simulations and extrapolating low-Mach-number results to higher Mach number.

\subsection{Circumplanetary disks}
Circumplanetary disks are different from quiescent CV disks mainly in that their Mach number tends to be much smaller,
especially at their outer edges, by a factor $\sim r_{\rm H}/a=(M_{\rm planet}/3M_{\rm star})^{1/3}$, in which $a$ is
the planet's orbital semimajor axis and $r_{\rm H}$ its Hill radius.
As a result, the excitation can still be relatively large even if the ILR is far outside the disk.

It is shown in all our models that increasing the disk size or density tends to increase the rate of accretion and angular momentum extraction. Therefore, there always exists an equilibrium state where the accretion onto the planet and angular momentum extraction balance the mass and angular momentum input from the accretion onto the disk.\footnote{For CV disks, such equilibrium in principle also exists, but outbursts are triggered before this equilibrium is reached.} This suggests that accretion is mainly limited by the mass input from the accretion stream, rather than the ability to remove excessive angular momentum from the disk.
\section{Summary and conclusion}\label{sec:summary}
In this paper we investigate the linear excitation of spiral density waves by tidal perturbation in CV and circumplanetary disks,
focusing on the excitation mechanism and the amplitude scaling with respect to the Mach number $\mathcal M$ of the disk when the inner Lindblad resonance (ILR) lies outside the disk.
We summarize our main results below, and discuss their implications.

\subsection{Wave excitation mechanisms and amplitude scaling}
The strength of wave excitation can be characterized by the dimensionless wave amplitude $|A_{\rm in}|=(F_{\rm in}/F_0)^{1/2}$, with the unit angular-momentum flux $F_0$ given in \eqref{eq:F0def}.

When the ILR lies inside the disk, wave excitation is dominated by the ILR and the dimensionless wave amplitude is
$|A_{\rm in}|\sim 1$, which is independent of $\mathcal M$ and insensitive to different disk models
(\S\ref{subsec:ILR_inside} and \S\ref{subsec:ILR}). This agrees with the result of \citet{GT79}.

When the ILR lies far outside the disk, there are several possible linear excitation mechanisms, and the scaling of
$|A_{\rm in}|$ with respect to $\mathcal M$ depends sensitively on the disk profile.
\begin{itemize}
\item When the disk contains no discontinuity or resonance, the excitation tends to be exponentially small, with
  $\log|A_{\rm in}| \sim -\mathcal M$ (\S\ref{subsec:ILR_outside}).
\item The disk may contain resonances other than the ILR, namely acoustic-cutoff resonance (ACR, \S\ref{subsec:ACR}) and
  stratification-shifted acoustic-cutoff resonance (SACR, \S\ref{subsec:SACR}). Both ACR and SACR require
  $|d\ln\Sigma_0/dr|\sim\mathcal M/r$, suggesting that they should be located near the disk edge. The angular momentum
  flux and amplitude of the excited wave can be analytically estimated by a local model.  The strength of excitation by
  ACR and SACR depends sensitively on the disk profile.
  The amplitude tends to be small, mainly limited by the surface density at the resonance.
\item Near-discontinuities in the disk profile (significant change over a radial distance $\lesssim r/\mathcal M$) can
  also produce nontrivial excitation, with $|A_{\rm in}|\propto \mathcal M^{-3/2}$
  (\S\ref{sec:sharp_edge}).
\item Excitation by the impact of the stream on the disk edge scales with the rate at which mass is added to the disk
  rather than the mass already there, and is therefore likely less important than tidal excitation for driving
  accretion.
\item Excitation of shocks at the streamline-crossing radius may truncate CV disks in outburst but is likely less
  important than the linear mechanisms discussed here
  for exciting waves in quiescent disks if these extend only to the circularization radius of the incoming stream.
\end{itemize}

\subsection{Applications and discussion}
We discuss the applications of our results in the context of CV disks and circumplanetary disks in \S\ref{sec:application}.
\begin{itemize}
\item For quiescent CV disks in which MRI is suppressed, density waves might drive the accretion rate if the resonance
  (ACR or SACR) happens at sufficiently large density, or if the disk has a sharp edge at which the sound speed does not
  vanish.\footnote{This presumes that the waves reach the inner parts of the disk before completely damping, as the
   three-dimensional nonlinear simulations of \cite{JSZ16,JSZ17} suggest; however, they
    ignore vertical stratification of density and temperature.  Density waves will not go far in disks
    that are hottest near their midplanes, as might be expected of MRI-turbulent disks with high optical depth
    \citep{Lubow+Ogilvie98}.  But passively illuminated quiescent disks may have the opposite thermal stratification,
    which favors long-range propagation \citep{Ogilvie+Lubow99}.} Linear theory also predicts a steep dependence on the
  disk size that is qualitatively consistent with observations.
\item For circumplanetary disks, these results suggest that there may exist an equilibrium state where the angular
  momentum added by accretion onto the disk is balanced by the tidal torque. As a result, accretion of the planet is not
  limited by the ability to remove excessive angular momentum from the disk.
\end{itemize}

More quantitative statements cannot be made with confidence for the lack of knowledge of the disk structure, especially in
its outer parts.  Our results suggest that discontinuities or ACR/SACR resonances near the disk edge are necessary for
significant excitation, at least in quiescent disks that lie well inside their ILR and streamline-crossing radii.  But
observationally determining the properties of the disk edge is difficult.  Nevertheless, if density waves are at all
important for the angular momentum and accretion of quiescent CV disks, the waves must be driven predominantly by the
tidal potential of the companion rather than the impact of the stream from the L1 point.

Our results also suggest that one has to be cautious when simulating the problem, since prescribing a disk model that is
not fully realistic may cause the result to deviate by orders of magnitude, especially when the disk Mach number is
large.  Moreover, extrapolating from simulations performed at Mach numbers $\mathcal{M}\sim 10-20$ to real disks at
$\mathcal{M}\sim 100$ is problematic, since the Mach number scaling is not likely to be a simple power law, and can be
correctly determined only when both the equilibrium disk model and the main wave excitation mechanism are known.
Because of the sensitivity to Mach number the temperature structure of the outer disk is just as important as its
surface-density profile (unless the disk extends to the ILR).  Simplified (e.g. isothermal) equations of state may give
misleading results.

\section*{Acknowledgements}

We thank Wendy Ju,  J.C.B. Papaloizou, Jim Stone, and Zhaohuan Zhu for useful discussions.

\bibliographystyle{mnras}
\bibliography{bib}

\newpage
\onecolumn
\appendix
\section{Validity of ignoring disk self gravity}\label{appendix:self_gravity}
Consider the gravitational potential perturbation $\varphi$ due to the density perturbation in the disk. Assume that the disk is 2D (i.e. zero thickness in $z$), we have
\eq{
\nabla^2\varphi = 4\pi G\Sigma\delta(z).\label{poisson}
}
For $z\neq 0$, the WKB solution of $\varphi$ is \cite{GT79}
\eq{
\varphi(r,z) = \Phi(r)\exp\left(-|h(r)z|+i\int^rh(s)ds\right),
}
with $\Phi(r)$ and $h(r)$ being real functions. Physically, $h(r)$ is the WKBJ frequency. Note that we have dropped the $\phi$ and $t$ dependence $\exp[im(\phi-\omega t)]$ as we do for other variables. Now \eqref{poisson} at $z=0$ becomes
\eq{
-2|h|\Phi\exp\left(i\int^rh(s)ds\right) = 4\pi G\Sigma.
}
To match the phase, for the most part of the disk we have
\eq{
h(r)\sim k(r)\sim c^{-1}.
}
Ignoring disk self gravity is a good approximation when $|\varphi(r,z=0)/K(r)|\ll 1$. Since
\eq{
\left|\frac{\varphi(r,z=0)}{K(r)}\right| \sim \left|\frac{c\Sigma}{c^2\Sigma/\Sigma_0}\right| \sim \Sigma_d\mathcal M,
}
the condition under which our approximation holds become
\eq{
\Sigma_d\mathcal M\ll 1.
}
Note that $\Sigma_d$, when the disk is not too small compared to the Roche lobe, is of the order of the disk to primary mass ratio.
This condition is satisfied for typical CV disks in quiescence and circumplanetary disks, therefore ignoring disk self gravity is a valid approximation for most part of the disk.

One caveat is that in the analysis above we are considering only regions in the disk where the WKBJ approximation is applicable. However, as we show in the main text, density waves are mainly excited where the WKBJ approximation does not hold.
Still, based on the agreement between our result (ignoring self gravity) and that of \citet{GT79} (including self gravity) for wave excitation by ILR, it is likely that self gravity in the wave excitation region will not significantly affect any nontrivial excitation.

\section{Eigenvalues of the system}\label{appendix:eigenvalues}
When written in the scale variables $\mathbf y = [\sqrt{r\Sigma_0}\xi,\sqrt{r\Sigma_0(K+W_m)}]^T$he first order equations \eqref{eq:eom1} and \eqref{eq:eom2} become
\eq{
\frac{d\mathbf y}{dr} = \mathbf{M\cdot y} + \mathbf w,\label{eq:y_first_order}
}
where
\eq{
\renewcommand*{\arraystretch}{2.5}
\mathbf M = \begin{bmatrix}
-\left[\dfrac{2m\Omega}{r\sigma}+\dfrac 12\dfrac{d\ln(r\Sigma_0)}{dr}-N_0^2\left(\dfrac{dK_0}{dr}\right)^{-1}\right] & \dfrac{m^2}{r^2\sigma^2}-\dfrac{1}{c^2}\\
\sigma^2-\kappa^2-N_0^2 & +\left[\dfrac{2m\Omega}{r\sigma}+\dfrac 12\dfrac{d\ln(r\Sigma_0)}{dr}-N_0^2\left(\dfrac{dK_0}{dr}\right)^{-1}\right]\\
\end{bmatrix},
}
\eq{
\renewcommand*{\arraystretch}{2.5}
\mathbf w = \begin{bmatrix}
\dfrac{1}{c^2}\sqrt{r\Sigma_0}W_m\\
N_0^2\left(\dfrac{dK_0}{dr}\right)^{-1}\sqrt{r\Sigma_0}W_m
\end{bmatrix}.
}
Note that $\mathbf M$ is traceless. The eigenvalues of the system, therefore, are both purely real or both purely imaginary. Note that this is not the unique choice of $\mathbf y$ that gives a traceless $\mathbf M$. However, any different scaling will make $\mathbf M$ directly depend on $\Sigma_0$, while here $\mathbf M$ depends only on $1/H\equiv -d\ln\Sigma_0/dr$.

The eigenvalue is $\pm ik_0$ with
\eq{
k_0^2 =& \frac{\sigma^2-\kappa^2-N_0^2}{c^2}\left(1-\frac{m^2c^2}{r^2\sigma^2}\right) - \left[\dfrac{2m\Omega}{r\sigma}+\dfrac 12\dfrac{d\ln(r\Sigma_0)}{dr}-N_0^2\left(\dfrac{dK_0}{dr}\right)^{-1}\right]^2\\
\approx & \frac{\sigma^2-\kappa^2-N_0^2}{c^2} - \left[\frac{1}{2H} + N_0^2\left(\dfrac{dK_0}{dr}\right)^{-1}\right]^2.
}
In the second line, all terms that are always $\lesssim O(1)$ are dropped. This gives the eigenvalue in \eqref{eq:k}.
\section{The second order equation and the local model for resonances}\label{appendix:local_model}
\subsection{The second order equation}\label{appendix:local_model1}
The first order equation \eqref{eq:y_first_order} can be rewritten as a first order equation in $y_1$ in the form of \eqref{eq:eom_y1}, with
\begin{subequations}\label{eq:coeffs}
\begin{align}
p(r) &= -\frac{d\ln |M_{12}|}{dr} \approx \frac{d\ln c^2}{dr},\\
k^2(r) &= M_{11}\left(\frac{d\ln M_{12}}{dr}-\frac{d\ln M_{11}}{dr}\right) + k_0^2\approx k_0^2,\\
\label{eq:fdef-full}
f(r) &= M_{12}\frac{d(M_{12}^{-1}w_1)}{dr} + M_{11}w_1 + M_{12}w_2\nonumber\\
&= \frac{1}{c^2}\sqrt{r\Sigma_0}\left(\frac{dW_2}{dr} - \frac{2m\Omega}{r\sigma}W_2\right) - 
\frac{d\ln(1-m^2c^2/r^2\sigma^2)}{dr}\frac{1}{c^2}\sqrt{r\Sigma_0}W_2 +
  \frac{m^2}{r^2\sigma^2}N_0^2\left(\frac{dK_0}{dr}\right)^{-1}\sqrt{r\Sigma_0}W_2
\nonumber\\
&\approx \frac{1}{c^2}\sqrt{r\Sigma_0}\left(\frac{dW_2}{dr} - \frac{2m\Omega}{r\sigma}W_2\right).
\end{align}
\end{subequations}
The approximate final equalities in each of eqs.~\eqref{eq:coeffs}, and the final form of $k_0^2$ in
eq.~\eqref{eq:k_approx}, are accurate when $\mathcal M\gg 1$ and $\mathcal M^2/r\gg
1/H, N_0^2(dK_0/dr)^{-1}$.

\subsection{Obtaining the local model for resonances}\label{appendix:local_model2}

Near a resonance (i.e. a zero of $k_0^2$), the second-order equation can be further approximated as
\begin{equation}\label{eq:airy_nonlocal}
  \frac{d^2y_1}{dx^2}-xy_1 = \lambda^2 f(r),
\end{equation}
with $x\equiv (r-r_{\rm res})/\lambda$ and $\lambda$ the lengthscale \eqref{eq:lambda}, provided that
\begin{equation}\label{eq:slow_varying_condition}
\lambda\ll r,\quad\left|\frac{d\ln M_{12}}{dr}\right| \ll \lambda^{-1}, \quad
\left|\frac{d}{dr}\left[M_{11}\left(\frac{d\ln M_{12}}{dr}-\frac{d\ln M_{11}}{dr}\right)\right] \right| \ll \lambda^{-3},
\end{equation}
A sufficient condition for \eqref{eq:slow_varying_condition} to hold is that
$c^2,H,N_0^2,dK_0/dr$ (and $\sigma^2-\kappa^2$ when the resonance is not the ILR) are effectively constant
over distances $\sim\lambda\ll r$.  

Note, however, that we do not require $\Sigma_0$ to be slowly varying:
indeed, its scale length $H$ may be comparable to $\lambda$ at an ACR or SACR, provided only that $H$ itself varies slowly.
The variation of $f(r)$ in eq.~\eqref{eq:airy_nonlocal} is then dominated by the factor \(\sqrt{\Sigma_0}\) that it
contains.  Defining $\alpha\equiv\lambda/2H$, we approximate eq.~\eqref{eq:airy_nonlocal} still further by
putting
\begin{equation}
  \label{eq:alpha_def}
  f(x)\approx f(r_{\rm res}) e^{-\alpha x},\qquad \alpha\equiv \lambda/2H\,,
\end{equation}
which gives eq.~\eqref{eq:airy}.

\subsection{Amplitude of the ingoing wave}\label{appendix:solve_lm2}

At large negative $x$, the solution to eq.~\eqref{eq:airy} should consist of an ingoing wave 
that satisfies the homogeneous Airy equation---plus a non-wavelike
part that is in phase with the local forcing term
\begin{equation}\label{eq:airy-expect}
  y_1(x)\approx A_{\rm in}[\mathrm{Ai}(x)+i\mathrm{Bi}(x)]\ +\ 
(-x)^{-1}\lambda^2f(r_{\rm res})e^{-\alpha x}\quad x\to-\infty.
\end{equation}
The angular-momentum flux is proportional to the square of the first (wavelike) term, but the second term
is much larger because of the exponential factor.  In the forbidden zone, $y_1\to 0$ as $x\to+\infty$.

To solve \eqref{eq:airy} for the coefficient $A_{\rm in}$, we first adopt a new dependent variable \(u(x)= e^{\alpha
  x}y(x)\), which satisfies
\begin{equation}\label{eq:uode}
  \frac{d^2u}{dx^2} -2\alpha \frac{du}{dx} + (\alpha^2-x)u(x) = \lambda^2f(r_{\rm res}).
\end{equation}
The solution for $u(x)$ should tend to zero in both directions, so it will have a Fourier transform $\tilde u(q)$,
\begin{equation}
  \label{eq:FTu}
  u(x) = \frac{1}{2\pi}\int\limits_{-\infty}^\infty e^{iqx}\tilde u(q)\,dq\,.
\end{equation}
The transform $\tilde u(q)$  satisfies the transform of eq.~\eqref{eq:uode},
\begin{equation}
  \label{eq:FTeqn}
  \frac{d\tilde u}{dq} + (-iq^2+2\alpha q + i\alpha^2)\tilde u = 2\pi i\delta(q)\lambda^2f(r_{\rm res})\,,
\end{equation}
with initial condition $\tilde u(q)=0$ at $q<0$, since the phase of
$u(x)$ should increase with $x$ to have negative radial group velocity:
\begin{equation}
  \label{eq:uintrep}
  u(x) = i\lambda^2f(r_{\rm res})\int\limits_0^\infty \exp\left[\tfrac{1}{3}iq^3-\alpha q^2+iq(x-\alpha^2)\right]\,dq.
\end{equation}
This strongly resembles the standard integral representations for the Airy functions, with convergence for all
$x$ thanks to the factor $\exp(-\alpha q^2)$ in the integrand.  For large negative $x$, two contributions dominate
the integral.  One of these comes from the lower endpoint $q\approx 0$:
\begin{equation}
  \label{eq:unonwave}
  u_{\rm nonwave}(x)\approx (\alpha^2-x)^{-1}\lambda^2 f(r_{\rm res})\ + O(x^{-3})\,.
\end{equation}
This matches the expected second term on the right side of eq.~\eqref{eq:airy-expect} to leading order in $(-x)^{-1}$.
The remaining contribution comes from the vicinity of the steepest-descent point, i.e. the point $q=q_0$ such that
$\partial \psi(q,x)/\partial q=0$ if $\psi(q,x)$ represents the argument of the exponential in the
integral~\eqref{eq:uintrep}.\footnote{There are two roots, \(q_0=-i\alpha\pm\sqrt{-x}\), but only one of these is close to the
  positive real axis if $x\ll -1$, and neither if $x\gg +1$.}  The result of the steepest-descent calculation matches
the asymptotic expansion of \(\mathrm{Ai}(x)+i\mathrm{Bi}(x)\) to leading order, and the coefficient implies
\begin{equation}
  \label{eq:Ain-airy-result}
  A_{\rm in} = i\pi\lambda^2 f(r_{\rm res}) e^{-\alpha^3/3}\,.
\end{equation}

\section{Wave excitation in the polytropic disks: Analytic results}\label{appendix:edge}
When ILR lies outside the disk, the integrand of \eqref{eq:Ain} (and hence the torque density on the disk) oscillates rapidly.
Such integrals tend to be exponentially small when
the envelope of the oscillation tapers slowly and smoothly to zero at both limits of integration.
The integration \eqref{eq:Ain} ends abruptly at $r_{\max}$, however, and furthermore has a branch
point there if $n=(\gamma-1)^{-1}$ is not an integer, because
\(\mathcal{W}^{-1}(r)\propto\Sigma_0(r)\propto\sim (r_{\max}-r)^n\).
This raises the possibility that the disk edge makes a contribution to eq.~\eqref{eq:Ain} that decreases as
some power of \(\mathcal M_0\) rather than exponentially.
To address this possibility, we analyse eq.~\eqref{eq:eom_y1} in a simpler local approximation.

When written in terms of $\tilde K\equiv K+W_m$ as the dependent variable instead of $y_1$, eq.~\eqref{eq:eom_y1} becomes
\eq{
\frac{d^2}{dr^2}\tilde K + \tilde p \frac{d}{dr} \tilde K + \tilde k^2 \tilde K = \tilde f,
}
with
\eal{
\label {eq:ptilde}
\tilde p(r) =& \frac{d}{dr}\ln\left(\frac{\Sigma_0r}{|\sigma^2-\kappa^2|}\right), \\
\label {eq:ktilde}
\tilde k^2(r) =& - \frac{2m\Omega}{r\sigma}\left[\frac{d}{dr}\ln\left(\frac{\Sigma_0\Omega}{|\sigma^2-\kappa^2|}\right)\right] - \frac{m^2}{r^2} + \frac{\sigma^2-\kappa^2}{c^2},\\
\label {eq:ftilde}
\tilde f(r) =& \frac{\sigma^2-\kappa^2}{c^2}W_m.  } At the outer edge of the disk, $\Sigma_0$ and $c^2$ both go to zero
and $\tilde p,\tilde k^2,\tilde f$ all have simple poles. The residue of $\tilde k^2$ defines a length scale $\delta$,
\begin{equation}
  \label{eq:delta}
  \delta^{-1}\equiv n\left[\frac{\sigma^2-\kappa^2}{\eta GM_1/r^2}+\frac{2m\Omega}{r\sigma}\right]_{r=r_{\max}}\,.
\end{equation}
Note that \(\delta/r\sim \mathcal O(\eta)\sim \mathcal O(\mathcal{M}_0^{-2})\).  So for the purpose of
studying wave excitation at the disk edge when $\mathcal{M}_0\gg 1$, it is reasonable to discard all but the leading-order
behaviors (poles) of the functions \eqref{eq:ptilde}, \eqref{eq:ktilde}, and \eqref{eq:ftilde}, and
to adopt a scaled independent variable
\begin{equation}
  \label{eq:xdef}
  x\equiv \frac{r-r_{\max}}{\delta}\,.
\end{equation}
The resulting simplified equation is
\begin{equation}
  \label{eq:localODE}
  x\frac{d^2 \tilde K}{dx^2} + n\frac{d\tilde K}{dx} - \tilde K = \mathcal F_0 
\end{equation}
The constant \(\mathcal F_0=\delta n [r^2(\sigma^2-\kappa^2)W_m/\eta GM_1]_{r=r_{\max}}\)
differs from \(W_m(r_{\max})\) by a small \(O(\eta)\) correction.  

Equation \eqref{eq:localODE} has a regular singular point at $x=0$.  The change of variable \(x=-z^2/4\)
makes it
\begin{equation}
  \label{eq:withz}
\frac{d^2\tilde K}{dz^2} + \frac{2n-1}{z}\frac{d\tilde K}{dz} + \tilde K = -\mathcal F_0\,.
\end{equation}
Clearly $n=1/2$ is a particularly simple case because the homogeneous solutions are then sinusoids.  The regular
solution of eq.~\eqref{eq:localODE} has a convergent power series in integral powers of $x$ and hence even powers of
$z$, so we take \(\cos z\) as the ``regular'' solution of eq.~\eqref{eq:withz}.  The ingoing wave is\footnote{Interior
  to the ILR, negative radial group velocity corresponds to a phase that
  decreases inward from the disk edge, and therefore with increasing $z$.} \(\exp(-iz)\), and the
coefficient of this wave as $z\to\infty$ is formally \(-i\int_0^\infty \mathcal F_0\cos z\,dz\) [compare eq.~\eqref{eq:Ain}]. As it
stands, this integral is not convergent.  But $\mathcal F_0$ is a proxy for \(W_m(r)\), which (since $m\ne0$) tends to zero as
$r\to0$, corresponding to $z\to\infty$.  So we should think of $\mathcal F_0$ as a smooth slowly decreasing function
\(\mathcal F(z/\mathcal M_0)\); furthermore $\mathcal F(t)=\mathcal F(-t)$ because \(W_m\) is a regular function of
\(r-r_{\max}=-z^2\delta /4\).  So the amplitude of the ingoing wave is
\((-i/2)\mathcal M_0 \int_{-\infty}^\infty \mathcal F(t)\cos(\mathcal M_0t) \,dt\), which is exponentially small
at large $\mathcal M_0$ since $\mathcal F(t)$ is smooth.

Returning now to the general case, let $\nu\equiv n-1$ and set $\tilde K= z^{-\nu}w(z)$, so that
eq.~\eqref{eq:withz} becomes
\begin{equation}
  \label{eq:Besseleqn}
  \frac{d^2w}{dz^2} + \frac{1}{z}\frac{dw}{dz} + \left(1-\frac{\nu^2}{z^2}\right)w = -z^\nu \mathcal F(z/\mathcal{M}_0)\,.
\end{equation}
The homogeneous solutions of this last equation are Bessel functions of order $\nu$.  The solution
corresponding to the regular homogeneous solution of \eqref{eq:localODE} is $J_\nu(z)$, the ingoing
wave is \(H_\nu^{(2)}(z)=J_\nu(z)-iY_\nu(z)\),  and the amplitude of this wave in the particular solution as
$z\to\infty$ is
\begin{equation}
  \label{eq:An}
  A_{\rm in}(n) \approx \frac{i\pi}{2}\int\limits_0^\infty z^{\nu+1} J_{\nu}(z) \mathcal F(z/\mathcal{M}_0)\,dz\,,
\end{equation}
which has made use of the Wronskian \(\mathcal{W}[J_\nu,\,H_\nu^{(2)}]=-i\mathcal{W}[J_\nu,\,Y_\nu]=2i/\pi z\).  We
write ``$\approx$'' rather than ``='' because eq.~\eqref{eq:An} is based on the local approximation \eqref{eq:localODE}
rather than the exact LWE.  This last integral is again exponentially small, which can be seen as follows.
Set \(\mathcal F(t)=g(t^2)\) (since $\mathcal F$ is even in its argument) and suppose that $g(u)$ has an
inverse Laplace transform $\hat g(s)$, so that \( \mathcal F(t) = \int\limits_0^\infty \hat g(s) e^{-st^2}\,ds\).
Putting this into eq.~\eqref{eq:An}, reversing the order of integration, and invoking \citet[\S11.4.29]{AS72} yields
\eq{\label{eq:Analt}
A_{\rm in}(n) &\approx \frac{i\pi}{2}\int\limits_0^\infty\left(\frac{\mathcal{M}_0^2}{2s}\right)^{\nu+1}\, 
\exp\left(-\frac{\mathcal{M}_0^2}{4s}\right)\, \hat g(s)\,ds \\
& = \frac{i\pi}{2}\mathcal{M}_0^2\int\limits_0^\infty (2\sigma)^{-\nu-1} e^{-1/(4\sigma)}\hat g(\mathcal{M}_0^2\sigma)\,d\sigma.
}
Note finally that \(\hat g(s)\) decreases faster than any power of \(s\) as \(s\to\infty\) because \(g(u)=\mathcal F(\sqrt{u})\)
has a convergent Taylor series at \(u=0\) (\(r=r_{\max})\), so that
\begin{equation*}
  \frac{d^kg}{du^k}(0) = (-1)^k\int_0^\infty \hat g(s) s^k\,ds
\end{equation*}
must exist.  Therefore the integral \eqref{eq:Analt} decreases faster than any power of \(\mathcal{M}_0^{-2}\) as
\(\mathcal{M}_0\to\infty\).

To sum up, by inspecting the integral \eqref{eq:Ain} we find that the excitation will be exponentially small except when the thin region near outer edge of the disk (with characteristic size $\delta\sim\eta$) makes a nontrivial contribution.
Then, we illustrated (using the local model discussed above) that the contribution from this region decreases faster than any power of \(\mathcal{M}_0^{-2}\) as
\(\mathcal{M}_0\to\infty\).
As the numerical result in \S\ref{subsec:ILR_outside} shows, this is indeed the case and the amplitude decreases exponentially as $\mathcal{M}_0$ increases.

\end{document}